\documentclass[twocolumn,showpacs,preprintnumbers,amsmath,amssymb]{revtex4}

\usepackage{graphicx}
\usepackage{dcolumn}
\usepackage{bm}

\begin{document}
\title{Dynamical properties across a quantum phase transition in the Lipkin-Meshkov-Glick model }

\author{Paolo Solinas}
\author{Pedro Ribeiro}
\author{R\'emy Mosseri}
\affiliation{Laboratoire de Physique Th\'eorique de la Mati\`ere Condens\'ee, CNRS UMR 7600,
Universit\'e Pierre et Marie Curie, 4 Place Jussieu, 75252 Paris Cedex 05, France}

\begin{abstract}
It is of high interest, in the context of Adiabatic Quantum Computation, to better understand the complex dynamics of a quantum system subject to a time-dependent Hamiltonian, when driven across a quantum phase transition. 
We present here such a study  in the Lipkin-Meshkov-Glick (LMG) model with one variable parameter.
We first display numerical results on the dynamical evolution across the LMG quantum phase transition, which clearly shows a pronounced effect of the spectral avoided level crossings.  We then derive a phenomenological (classical) transition model, which already shows some closeness to the numerical results. Finally, we show how a simplified quantum transition model can be built which strongly improve the classical approach, and shed light on the physical processes involved in the whole LMG quantum evolution. From our results, we argue that the commonly used description in term of Landau-Zener transitions is not appropriate for our model.

\end{abstract}


\maketitle

\section{Introduction}

Under a continuous change of the Hamiltonian parameters, a quantum system, initially in its ground state, can undergo transitions to excited states. This point was already studied in the early days of quantum mechanics, with the celebrated analysis of the two level case by Landau\cite{landau32} and Zener \cite{zener32}.
Here we first display some numerical results on the dynamical evolution across the Lipkin-Meshkov-Glick (LMG) quantum phase transition with two very different pattern whenever the critical point is reached from both sides of the quantum phase transition (QPT). We then write down a phenomenological (classical) transition model (paragraph III), which already shows some closeness to the numerical results. Finally, we show how a simplified quantum transition model (paragraph IV) can be built which strongly improve the classical approach, and shed light on the physical processes involved in the whole LMG quantum evolution.
This question is again under a strong focus following the raise of the so-called ``Adiabatic Quantum Computation'' (AQC) \cite{AQC_proposal} approach to quantum computation. The basic idea here is to initially prepare a quantum system in an (easy to prepare) ground state of a known Hamiltonian $H_0$ and to encode the answer of a given computational problem in the ground state of a final (or ``Problem'') Hamiltonian $H_P$. The simplest way to implement AQC is to take, as time-dependent Hamiltonian $H(t)$, a linear interpolation between $H_0$ and $H_P$,
%
%
\begin{equation}
 H(t) = (1-\frac{t}{T}) H_0 + \frac{t}{T} H_P
 \label{eq:ad_ham}
\end{equation}
%
%
with $t$ the physical time, and $T$ the total evolution time. In the following, we shall use $s=t/T$, with $s\in [0,1]$.

According to the quantum adiabatic theorem, if passing from the initial to the final Hamiltonian is done slowly enough, the system remains in the ground state, and therefore, at the end of the evolution, will be in a state that encodes the solution of the computational problem. 
 
A clear distinction between standard (gate-like) quantum computation (QC) and AQC approaches is the way algorithms {\it computational efficiency} is quantified. In the standard QC case, the complexity refers to the number of logical gates to be applied as a function of $n_q$, the number of qubits. In AQC, the computational efficiency is related to the value that $T$ must be given, such that the adiabatic approximation holds, roughly estimated as \cite{messiah}
%
%
\begin{equation}
 T \gg \Delta_{min}^{-2} 
\label{eq:ad_theorem}
\end{equation}
%
%
where $\Delta_{min}$ is the minimum value of the energy gap between the ground and the first excited states taken along the evolution. The way $\Delta_{min}$ (and therefore $T$) scales as function of $n_q$ leads then to the computational efficiency in the AQC case.

Despite these differences, AQC has been proved to be (computationally) equivalent to QC \cite{equivalence_QC_AQC,complexity_AQC}. Its main interest is nevertheless that it provides a direct physical implementation, and, to some respect, a different kind of robustness to errors and decoherence \cite{childs02}. 

The QC-AQC equivalence calls for a physical effect responsible of AQC failure for highly complex computational problems. Since adiabatic quantum evolution is strongly sensitive to energy differences from ground to excited states, quantum complexity is clearly expected when gaps vanish for some value of $s$, which is the usual signature of a quantum phase transition \cite{QPT}. Such situations with gaps closing exponentially with $n_q$ have indeed been numerically observed for computationally hard problems \cite{complexity_AQC, complexity_AQC2}.

One therefore faces the interesting, although expected, picture that it is not the ground state itself (for $s=1$) that characterizes the problem complexity, but the nature of the process, in parameter space (for $s < 1$).
This should be related to the already noticed strong relationship between AQC and Quantum Annealing problems, well studied in the past years in the field of complex system (e.g. spin glasses) \cite{santoro_tosatti}.

It is therefore of high interest to study, in the transition region, the dynamical and spectral properties in the lower part of the spectrum. In particular, one should not only focus on the first excited eigenstate coming close to the ground state, but in fact to a whole set of excited levels. This is in particular expected whenever the ground state nature is drastically changed, since its new decomposition is mainly weighted by the states for which avoided crossings appear along the adiabatic process. So the state dynamical evolution has to be thought as a complex transition cascade, rather than independent Landau-Zener (LZ) processes.
Therefore, despite its great success in other systems, the LZ theory, in particular its ability to calculate the transitions between quantum states, may not be used in the present AQC case, or be severely corrected \cite{LZ_time,corrections_LZ}. It should be stressed in addition that LZ theory is intrinsically non-adiabatic, which suggest, for the AQC slow evolution processes, to go back to more standard adiabatic analysis\cite{messiah}.

To get a better understanding of these processes, we propose here to study the dynamical properties across the quantum phase transition in the Lipkin-Meshkov-Glick model, a simple (solvable) model which exhibits some of the expected features of AQC Hamiltonians. The LMG model was introduced long ago in $1965$ to study shape transitions in nuclei \cite{Lipkin65}, and has been, since then, proposed to describe many systems ranging from interacting spin systems \cite{Botet83} to Bose-Einstein condensates \cite{Cirac98} or magnetic molecules such as ${\rm Mn}_{12}$ acetate \cite{Garanin98}. 

The LMG model describes a set of $N$ spins $\frac{1}{2}$ mutually interacting through a $XY$-like Hamiltonian and coupled to an external transverse magnetic field $h$. This Hamiltonian $H$ can thus be expressed in terms of the total spin operators $S_{\alpha}=\sum_{i=1}^N \sigma_{\alpha}^{i}/2$ where the $\sigma_{\alpha}$'s are the Pauli matrices:
%
%
\begin{equation}
\label{eq:hamiltonian}
H=-\frac{1}{N} \big(\gamma_x S_x^2 + \gamma_y S_y^2  \big) - h \: S_z,
\end{equation}

In the following, we only consider the maximum spin sector $S=N/2$, with $N$ even and $N+1$ levels.
Although many different methods have been used to study its excitation properties, the richness of the full spectrum has only be revealed quite recently by means of numerical diagonalizations \cite{Heiss05,Castanos06}, and then, at the thermodynamic limit, in an analytical form \cite{ribeiro_PRL}. Of interest here is the determination of the so-called ``exceptional points'' in the density of states, where the density of states is singular and the level separation vanish with $N$. These points gather, as $N$ tends to $\infty$, on a curve which we call here the ``critical gap curve'' (CGC).

Here, for sake of simplicity, we set $ \gamma_y=0$; it is clear that, up to a global reparametrization, the Hamiltonian only depends on the ratio between $\gamma_x$ and $h$. To express this Hamiltonian in the form given by equation (\ref{eq:ad_ham}), we write $ \gamma_x=s$, $h=1-s$, which leads to $  H_0 = - S_z $ and  $H_P=-S_x^2/N$

\begin{figure}[Ht]
  \begin{center}
   \includegraphics[height=5cm]{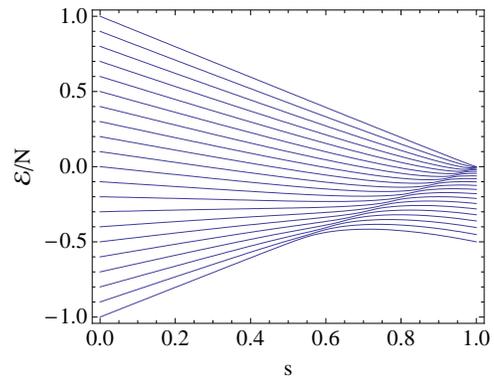}
    \caption{\label{fig:spectrum} Spectrum of the ``s''-dependant Lipkin-Meshkov-Glick model for $N=20$, with $21$ levels.  }
  \end{center}
\end{figure}

The energy levels as a function of $s$ are displayed in figure \ref{fig:spectrum} for $N=20$. 
At the thermodynamic limit, this system undergoes a second order quantum phase transition for $s=1/2$, whose effect is already visible with $N=20$, in terms of levels pinching.
The locus of avoided crossing levels appears very close to a straight line (exact CGC for infinite $N$) starting at the QPT for $s=0.5$, and $E/N= -0.5$, and reaching $E=0$ for $s=1$.

In later plots, we shall use the (normalized to one) integrated density of states $x$, in the range $[0,1]$, instead of the energy. The CGC still has a simple expression, $x_c(s)$,  which reads

\begin{equation}
\label{eq:mgc}
x_c(s) = 1-\frac{4}{\pi } \cot ^{-1}\left(\frac{\sqrt{s}+\sqrt{2  
s-1}}{\sqrt{1-s}}\right)-\frac{2}{\pi  s} \sqrt{(1-s) (2 s-1)}
\end{equation}

for $s \in [0.5,1]$

In the following, we first display some numerical results on the dynamical evolution across the LMG quantum phase transition (paragraph II),  with two very different pattern whenever the critical point is reached from one side and another of the QPT. We then write down a phenomenological (classical) transition model (paragraph III), which already shows some closeness to the numerical results. Finally, we show how a simplified quantum transition model (paragraph IV) can be built which strongly improves the classical approach, and shed light on the physical processes involved in the whole LMG quantum evolution

\section{Dynamical evolution : numerical results}

\subsection{Forward evolution}

We numerically solve the dynamical (arbitrarily called ``forward'') evolution, proceeding as follows. The initial state is the ground state corresponding to the $s=0$ Hamiltonian. The total evolution time $T$ is a multiple of a fixed time interval $\Delta T$, during which the Hamiltonian parameters are kept fixed and the quantum evolution is computed by mean of a standard second order discretization method. The final state, after $\Delta T$, serves as the initial state for the next step, with (slightly) varied Hamiltonian parameters. Therefore, the larger the $T$ value, the  smaller the effective Hamiltonian variations from one step to the next, and therefore the closer to an adiabatic evolution. Another parameter is $N$, the  system size. Increasing $N$ decreases the gaps, which eases the transitions to excited states. Computations are done here with $N=50$ and three $T$ values, corresponding qualitatively to fast ($T=1$), medium ($T=50$) and slow ($T=100$) evolutions.

\begin{figure*}[Ht]
  \begin{center}
   \includegraphics[height=5cm]{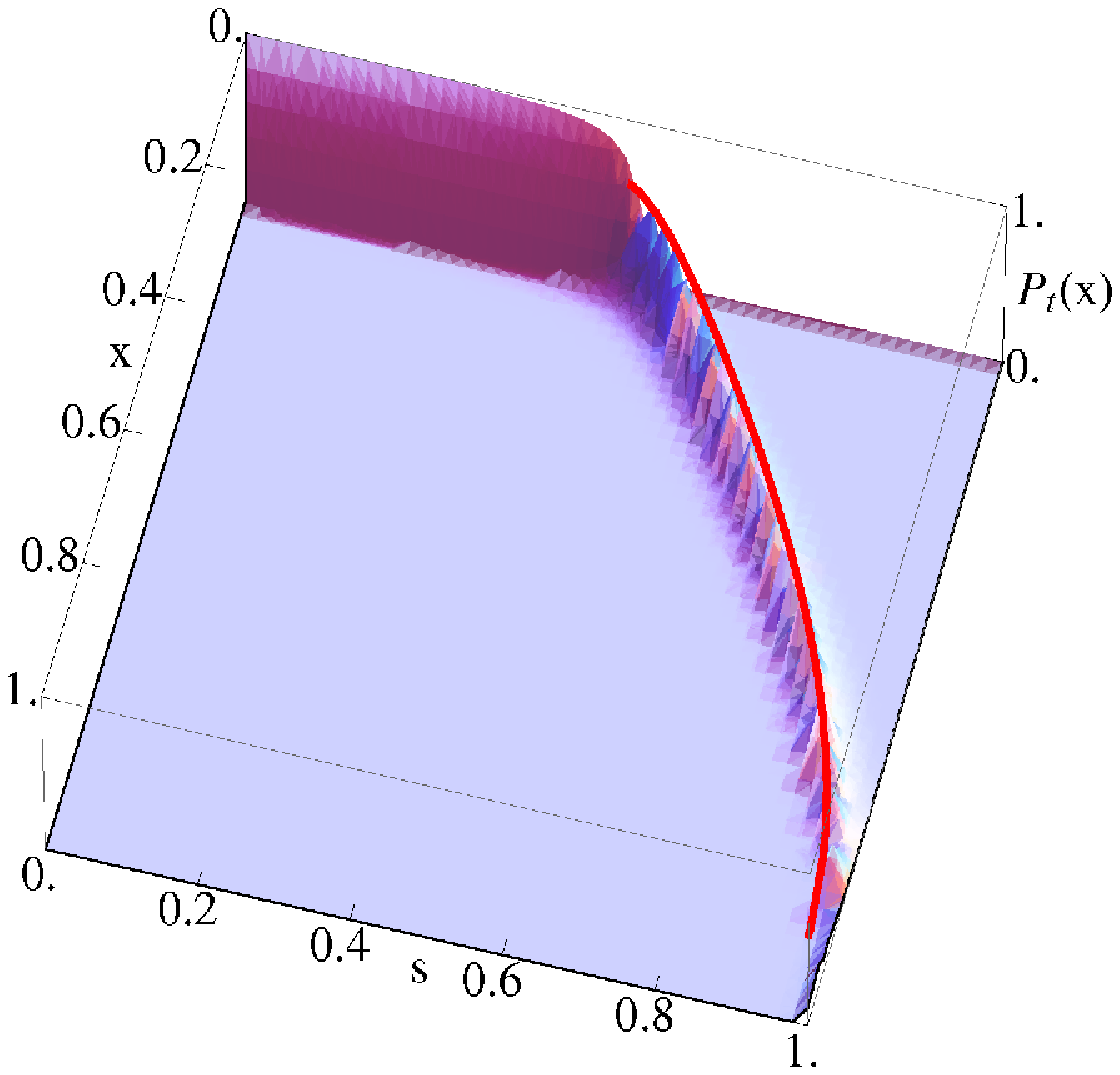}
   \includegraphics[height=5cm]{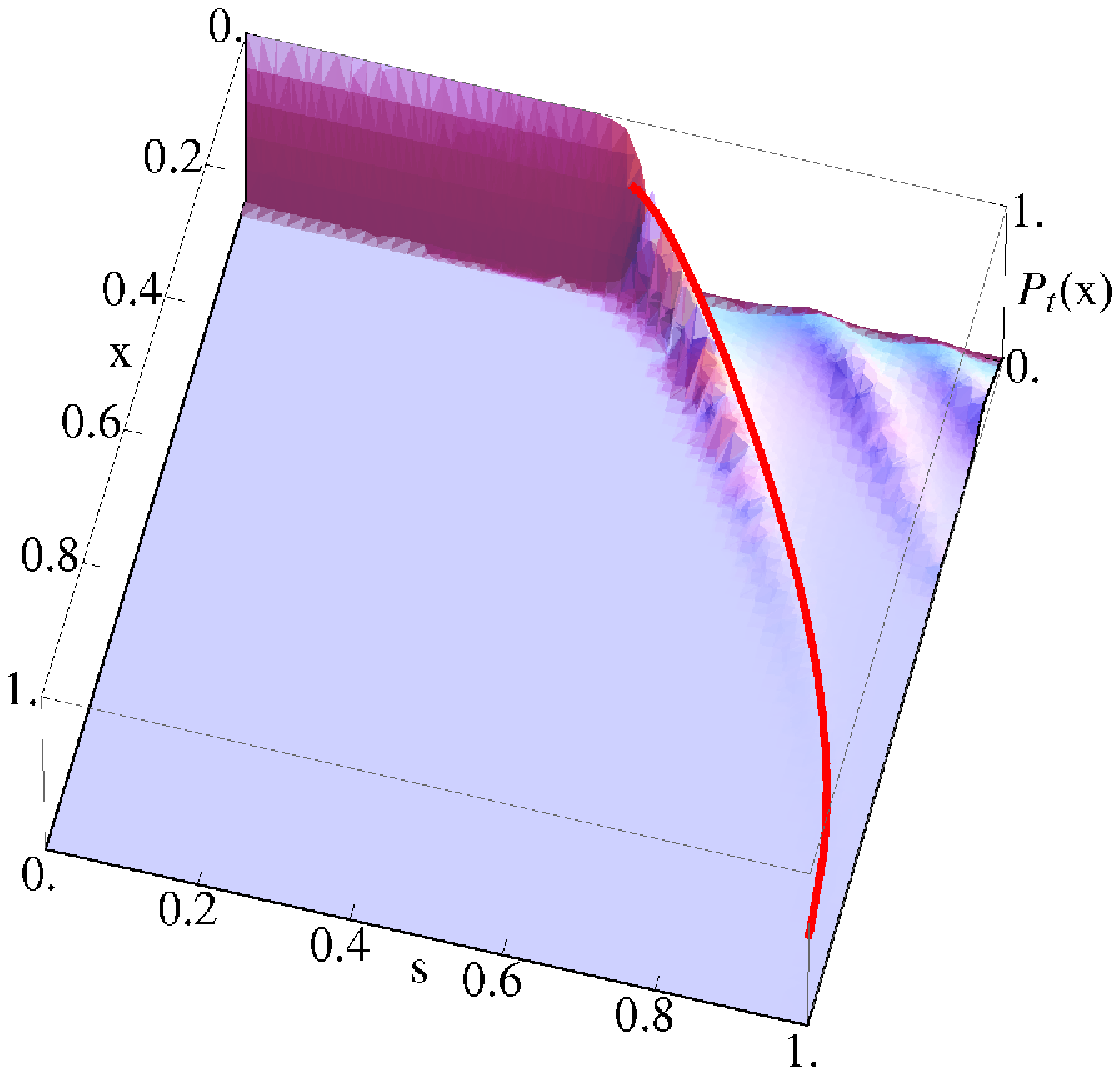} 
     \includegraphics[height=5cm]{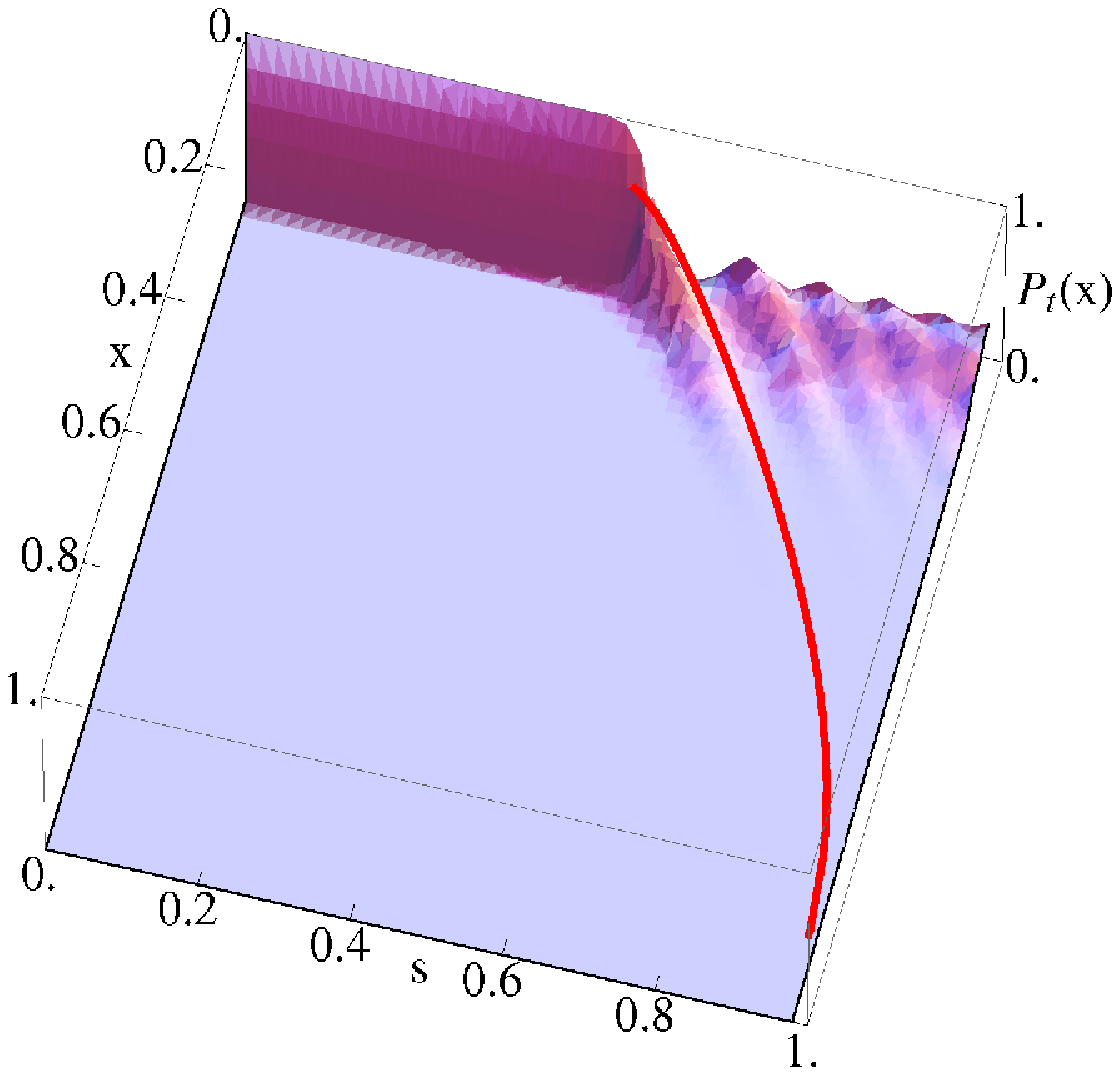} 
   \\
   \includegraphics[height=5cm]{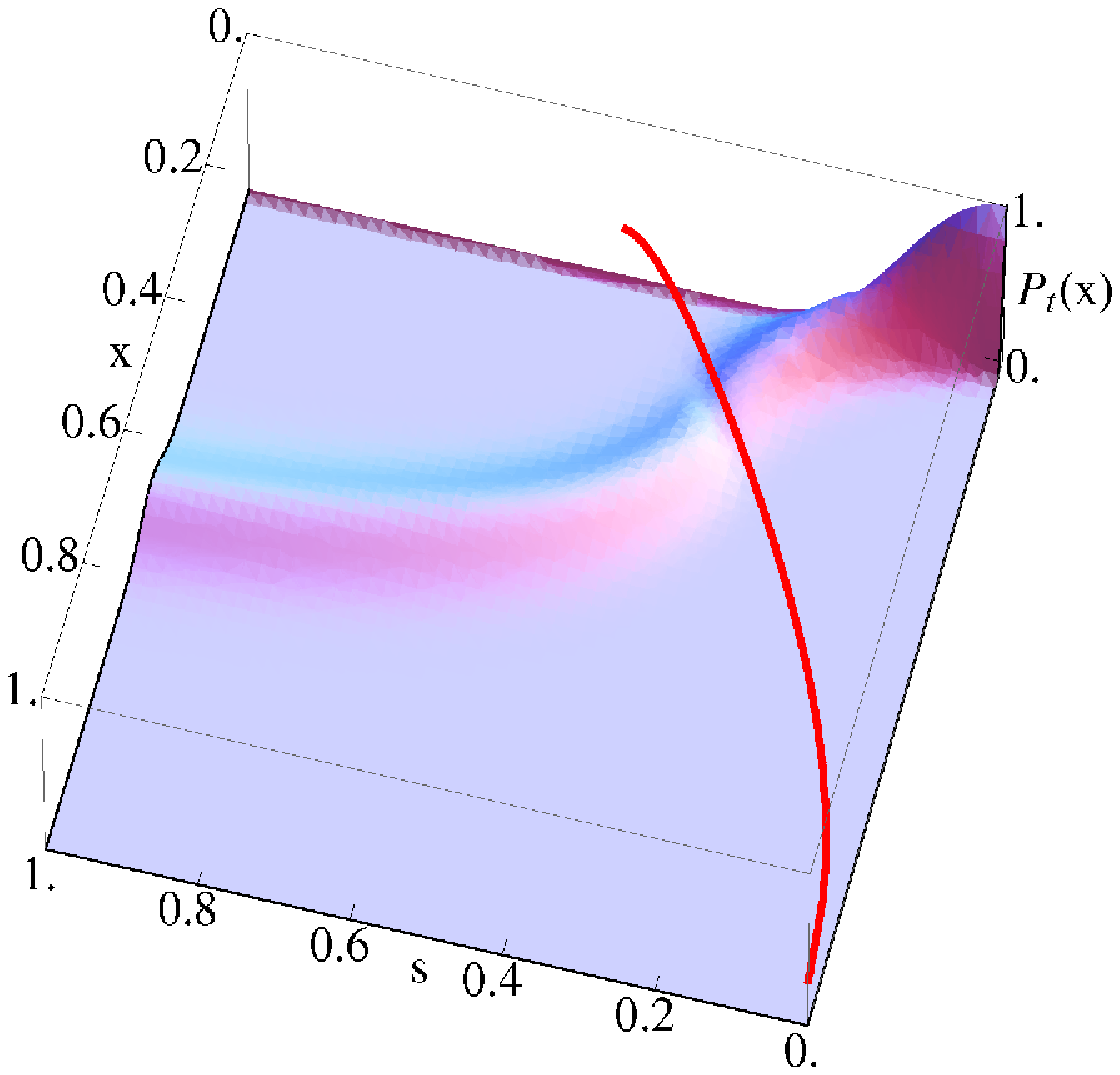}
   \includegraphics[height=5cm]{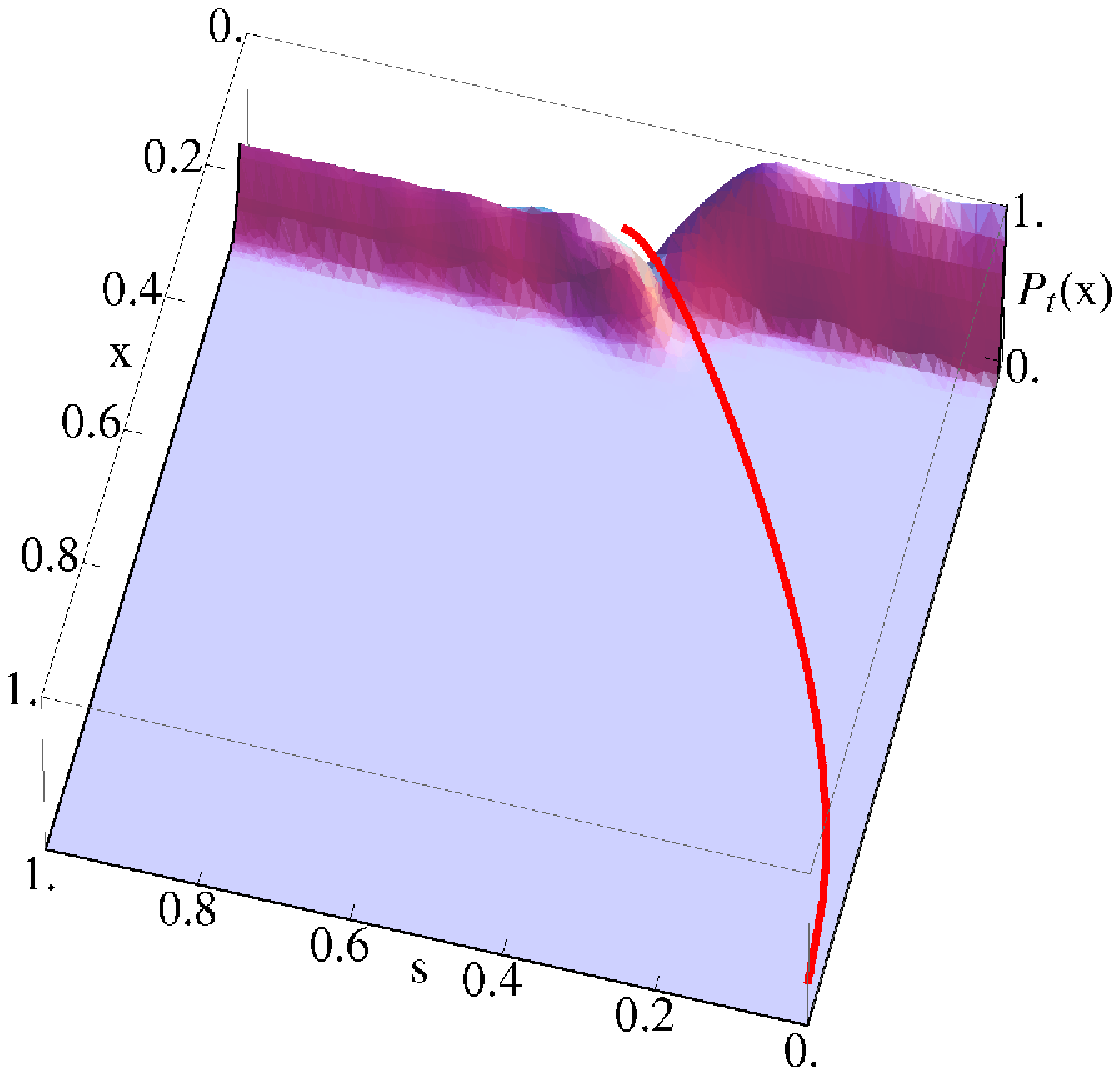}
    \includegraphics[height=5cm]{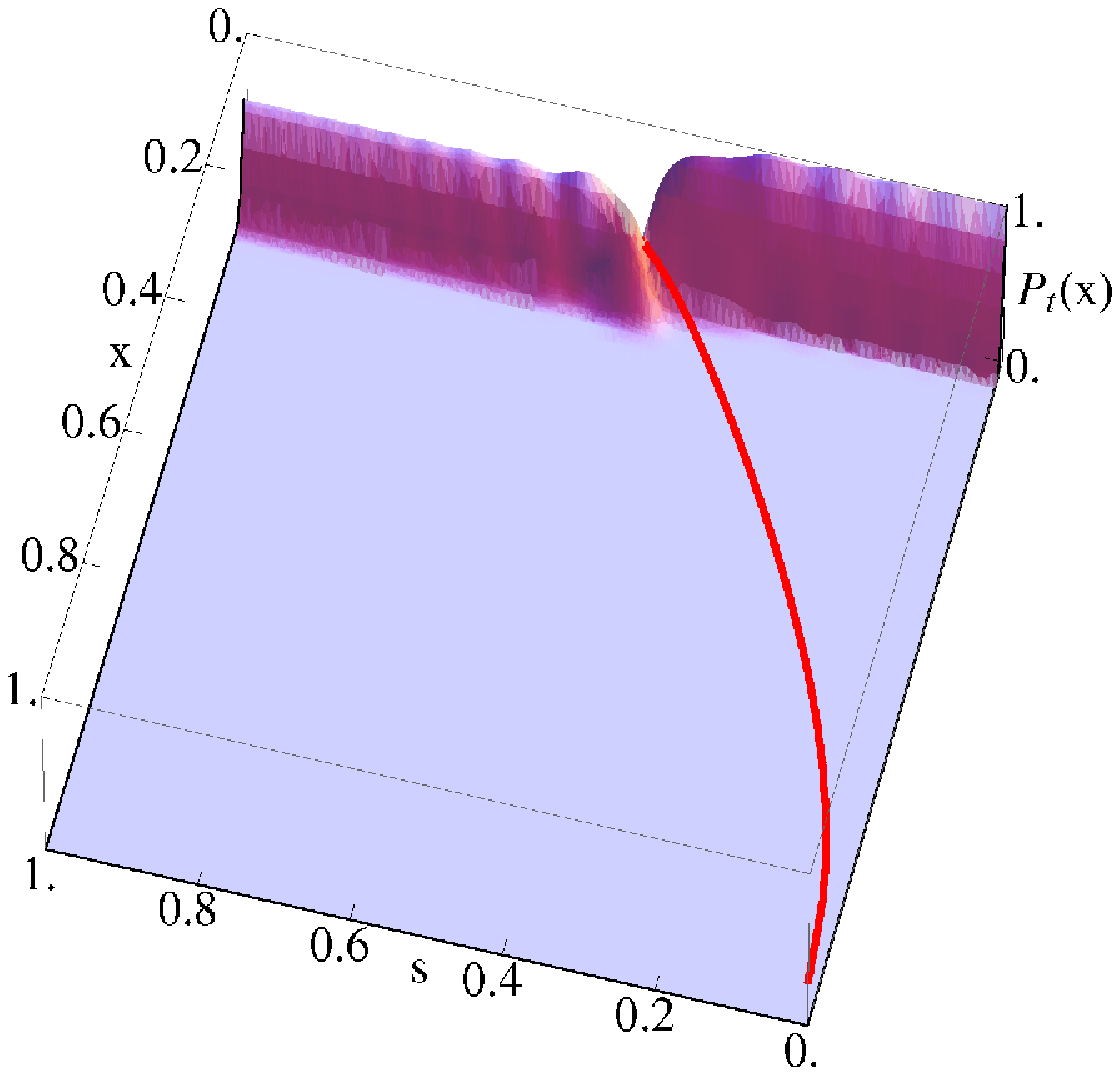}
 
    \caption{\label{fig:large_system_evolution} Forward and backward evolution of the level populations for a system of size $N=50$ for different the total evolution times $T$. The red line is the analytical CGC curve(see text). The three upper figures correspond to forward evolution : left, fast evolution ($T=1$); middle, intermediate speed ($T=50$); right, slower evolution ($T=100$).  The three lower figures correspond to the backward evolution, with the same speed ($T$ values) as the plots above }
  \end{center}
\end{figure*}

The levels occupancy, as a function of $s$, are displayed in Figure \ref{fig:large_system_evolution} (upper plots). 
Also shown is the CGC curve, in order to track the role of the gap closing phenomenon in the quantum evolution.

As can be clearly seen, a common feature of these evolutions is that the system almost remains in its ground state before reaching the quantum phase transition region. 

Then, not only do the quicker evolutions drive the system to excited states transitions, but this evolution is clearly controlled by the position of the avoided crossings, as marked by the critical gap curve.

As expected, for slower evolutions (larger $T$), the ground state is not completely depleted, its population  oscillates  with time (as seen on the figure) and eventually stabilizes (see for example Ref. \cite{LZ_time}).

\subsection{Backward Evolution}

The above observation that the whole spectrum influences the overall quantum state evolution, together with the fact that the LMG spectrum is far from being symmetrical (see figure  \ref{fig:spectrum}), leads to expect a qualitatively different time evolution whenever the system is driven backward, which reads

\begin{equation}
 H_{inv}(t) = s H_0 + (1-s) H_P
\end{equation}

The levels occupancy, displayed in Figure \ref{fig:large_system_evolution} (lower plots),  indeed  shows a very different pattern. A first explanation arises quite naturally : in the forward case, the system encounters the minimal energy gaps in an ordered sequence that allows the current wave function to spread in the spectrum, with a high probability to change its eigenstate decomposition along the avoided crossings. In the latter (backward) case, once the system encounters the first small gap, and possibly leaves the ground state, it never meets again avoided crossings situations, and therefore do not proceed significantly to higher energies.

In addition, the levels population displays other qualitative features which can be explained by looking to the spectrum. In the right part of the spectrum (Figure \ref{fig:spectrum}), which is first visited in the backward evolution, the gaps are smaller with respect to those in the left part. As a consequence, for the same $T$ value, transitions to excited states are more probable, and start occurring even before crossing the quantum phase transition (Figure \ref{fig:large_system_evolution} (lower left)).

Note finally (Figure \ref{fig:large_system_evolution}) that the states population is slightly less peaked (it displays a sudden small {\it anti-bump}) when the CGC curve is crossed. This is again a manifestation of the enhanced transition probability along this curve.

\section{A phenomenological  model for the quantum evolution} 

\begin{figure*}[Ht]
  \begin{center}
   \includegraphics[height=5cm]{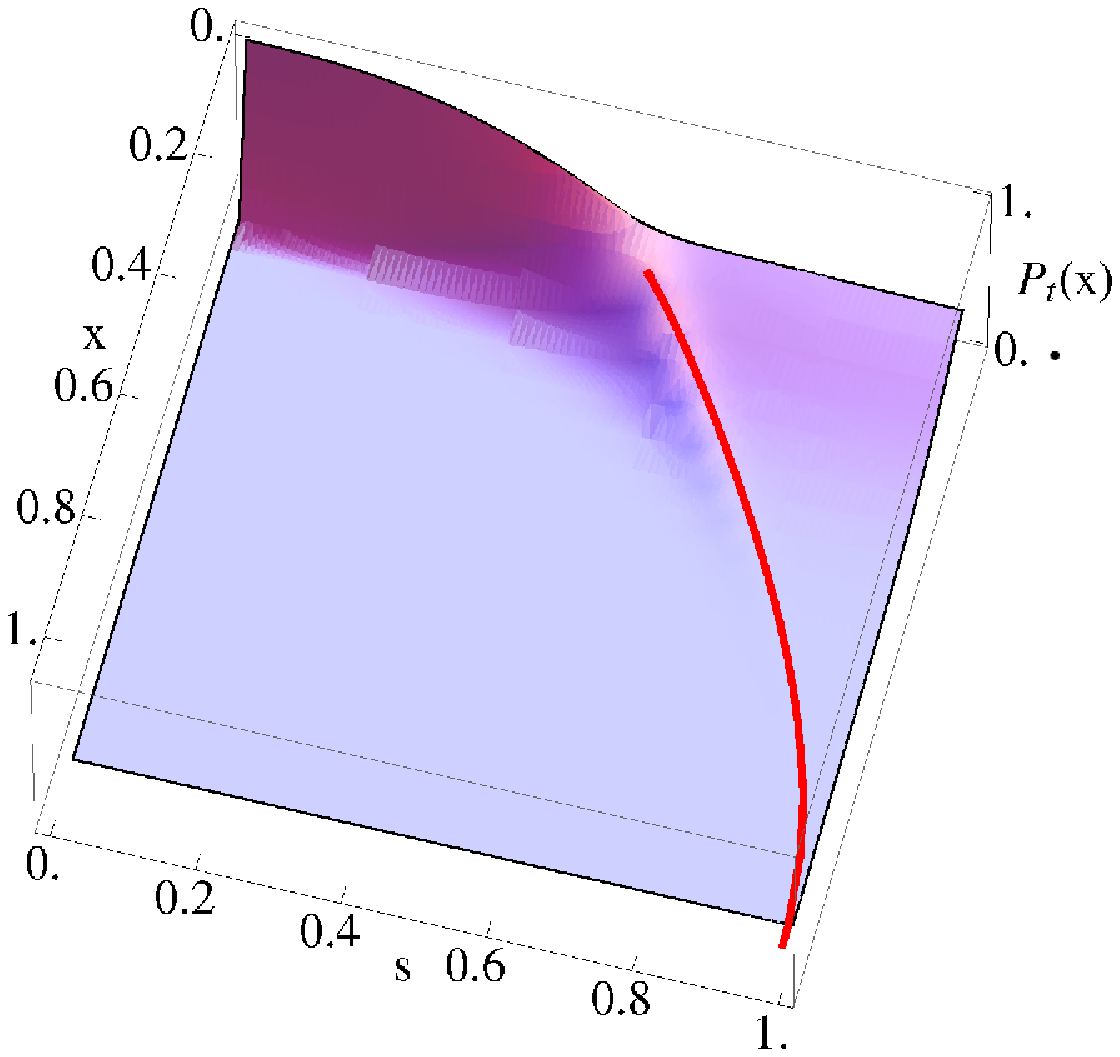}
   \includegraphics[height=5cm]{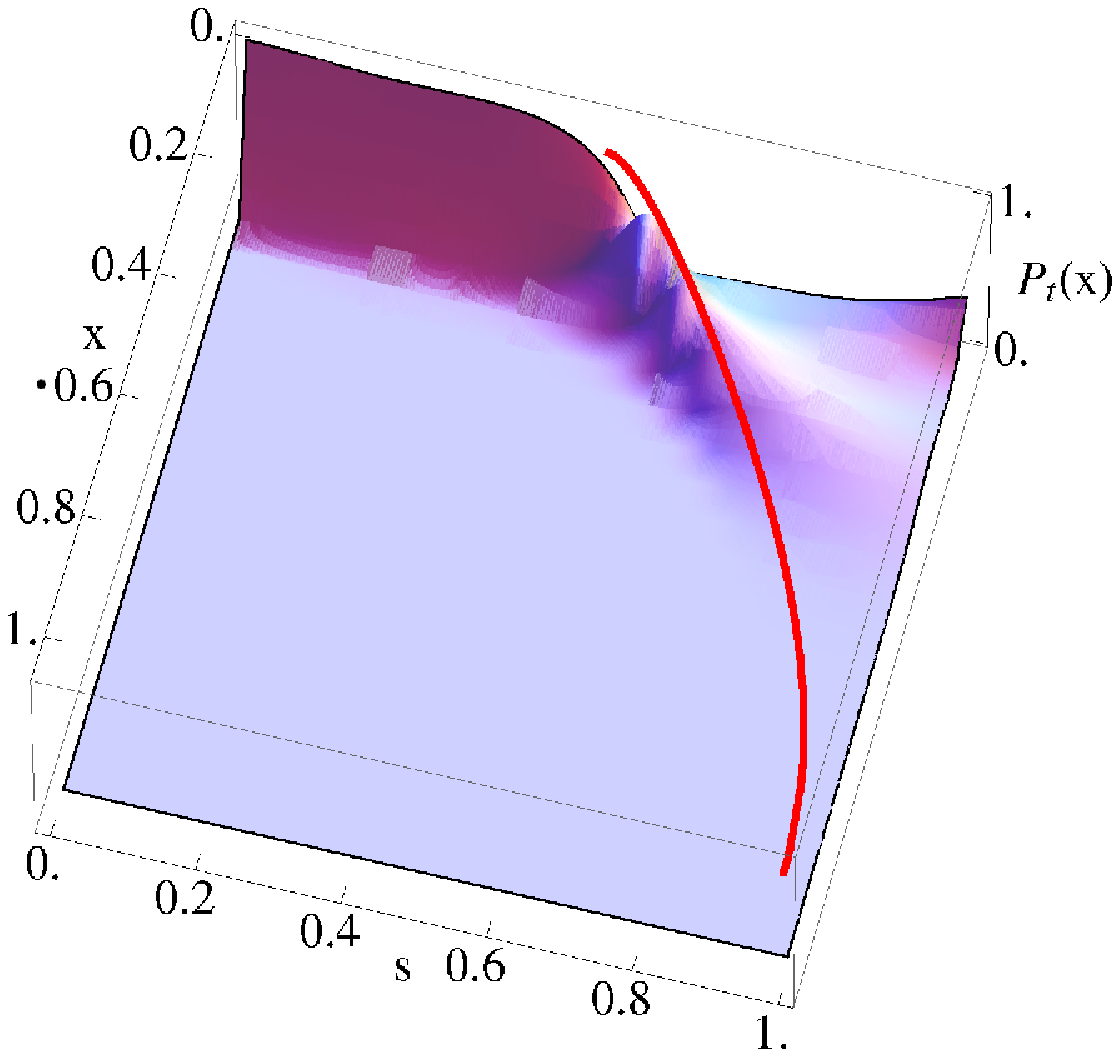} \\
   \includegraphics[height=5cm]{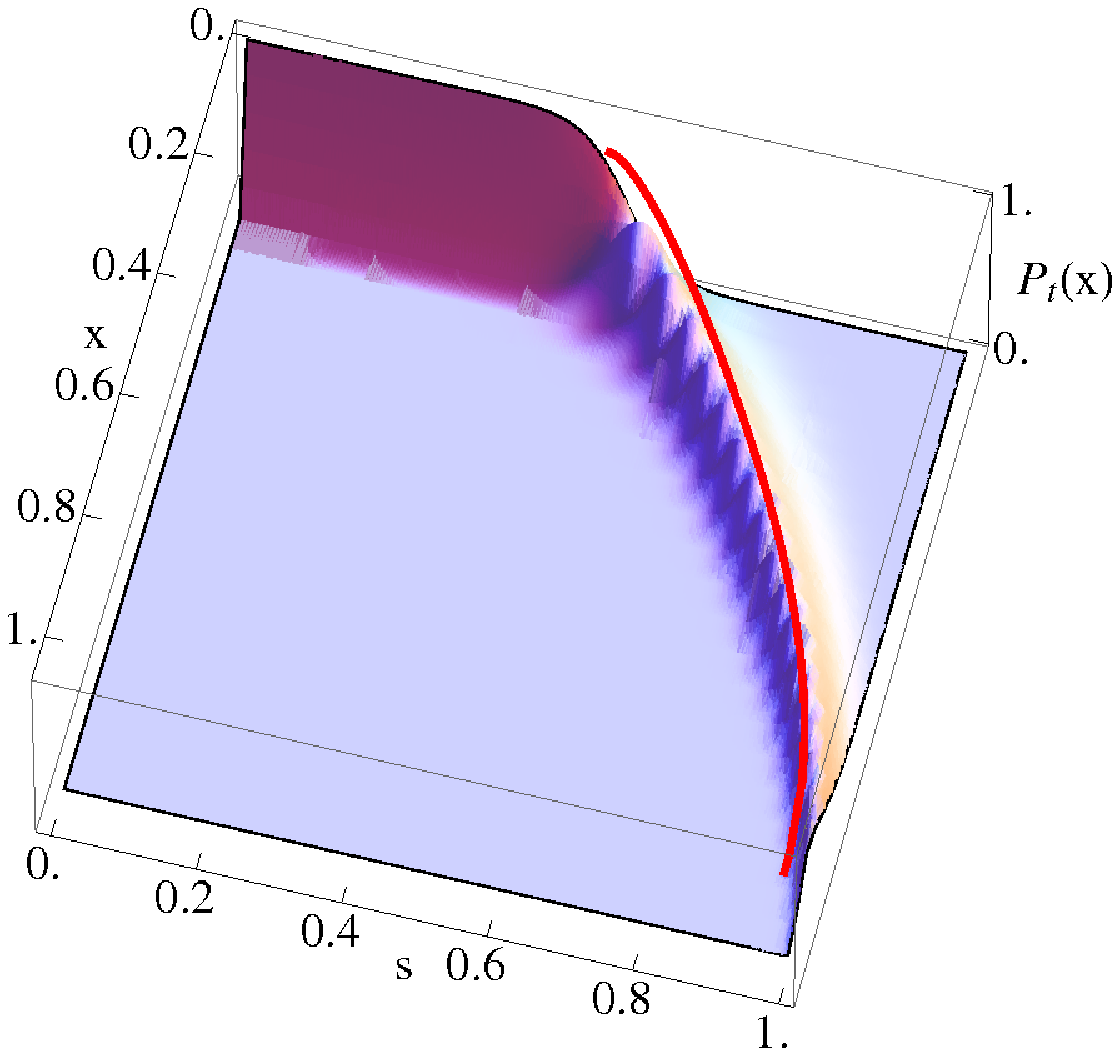}
   \includegraphics[height=5cm]{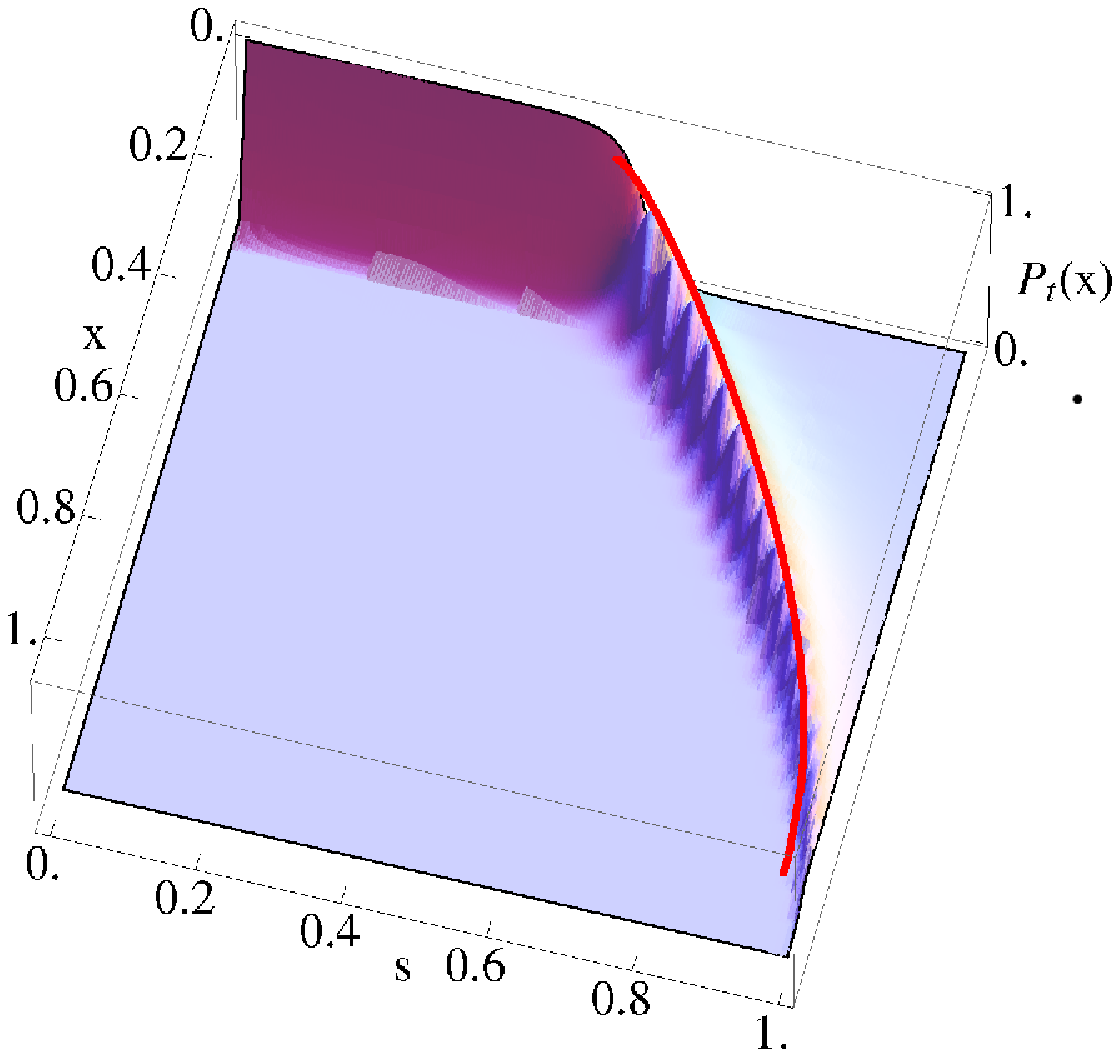}
    \caption{\label{fig:ME_direct_evolution} Levels population in the forward evolution case for a $20$ level system and $T=10$.  Phenomenological rate equation ($T~b=0.01$) (up left), effective quantum chain with constant matrix element (up right), effective chain with improved matrix elements (down left), full numerical evolution (down right).      }
  \end{center}
\end{figure*}

We shall be interested, in this chapter and the next one, in computing approximate values for the evolution of the probability distribution among the different states. 

We call $\{|\eta_i (s) \rangle \}$ the instantaneous eigenbasis  of the $s$-dependant Hamiltonian, $H(s) |\eta_i (s) \rangle = \epsilon_i(s) |\eta_i) (s) \rangle$, and write the current state$|\Phi(s)\rangle $, in this basis : $|\Phi(s)\rangle = \sum_i a_i(s) |\eta_i (s) \rangle$. 
We aim to compute the probability $P_i(s)=|\langle \eta_i (s)| \Phi(s)\rangle|^2$ for the system to be  in the $i$-th instantaneous eigenstate at time $s$.
We shall first suppose that transitions only occur, at the same rate, from an instantaneous eigenstate $|\eta_i \rangle$ toward states $|\eta_{i+1}\rangle$ and $|\langle\eta_{i-1}\rangle$. The transition rate matrix $\Gamma$ inherits a tridiagonal form, but with nevertheless $s-$dependant elements.
This leads to the following differential equation for the probability 

 \begin{eqnarray}
 \dot{P}_i(s) &=&  \Gamma_{i+1,i}(s) P_{i+1}(s)  + \Gamma_{i-1,i}(s) P_{i-1}(s) \nonumber \\ 
              & & -(\Gamma_{i,i+1}(s)+\Gamma_{i,i-1}(s)) P_i(s)  
\label{eq:classicalMQ}
 \end{eqnarray}

We choose a form for the rates $\Gamma_{i,i \pm 1}(s)$ which follows a generic adiabatic prescription

\begin{equation}
  \Gamma_{i,j}(s)= \frac{T~b}{(\Delta^{i,j}(s))^2}
	\label{eq:tr_rate}
\end{equation}
where $\Delta^{i,j}(s)$ is the instantaneous gap between levels $i$ and $j$, calculated from the spectrum, $T$ is the evolution time and $b$ a (adjustable) coupling parameter .

\begin{figure*}[ht]
  \begin{center}
   \includegraphics[height=5cm]{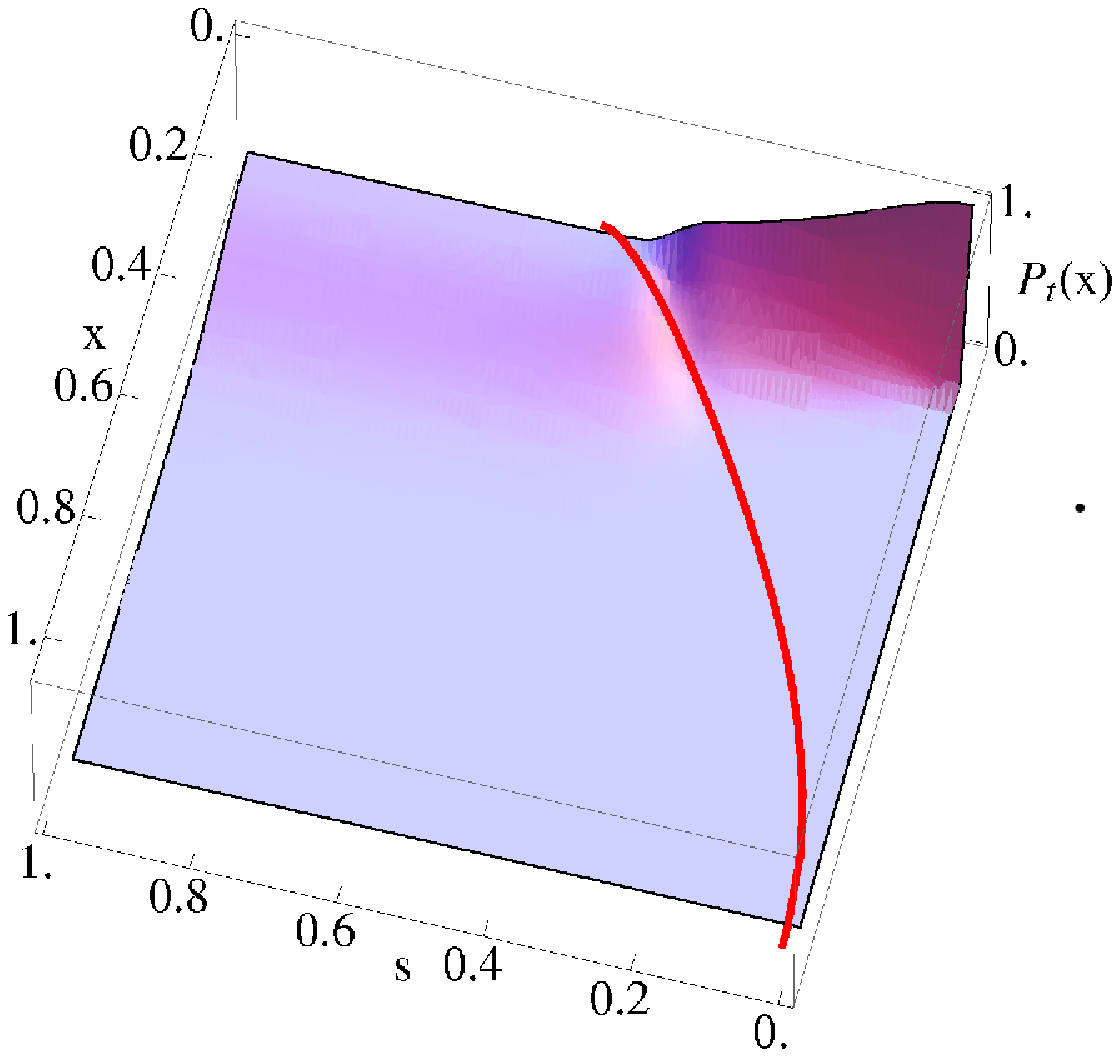}
   \includegraphics[height=5cm]{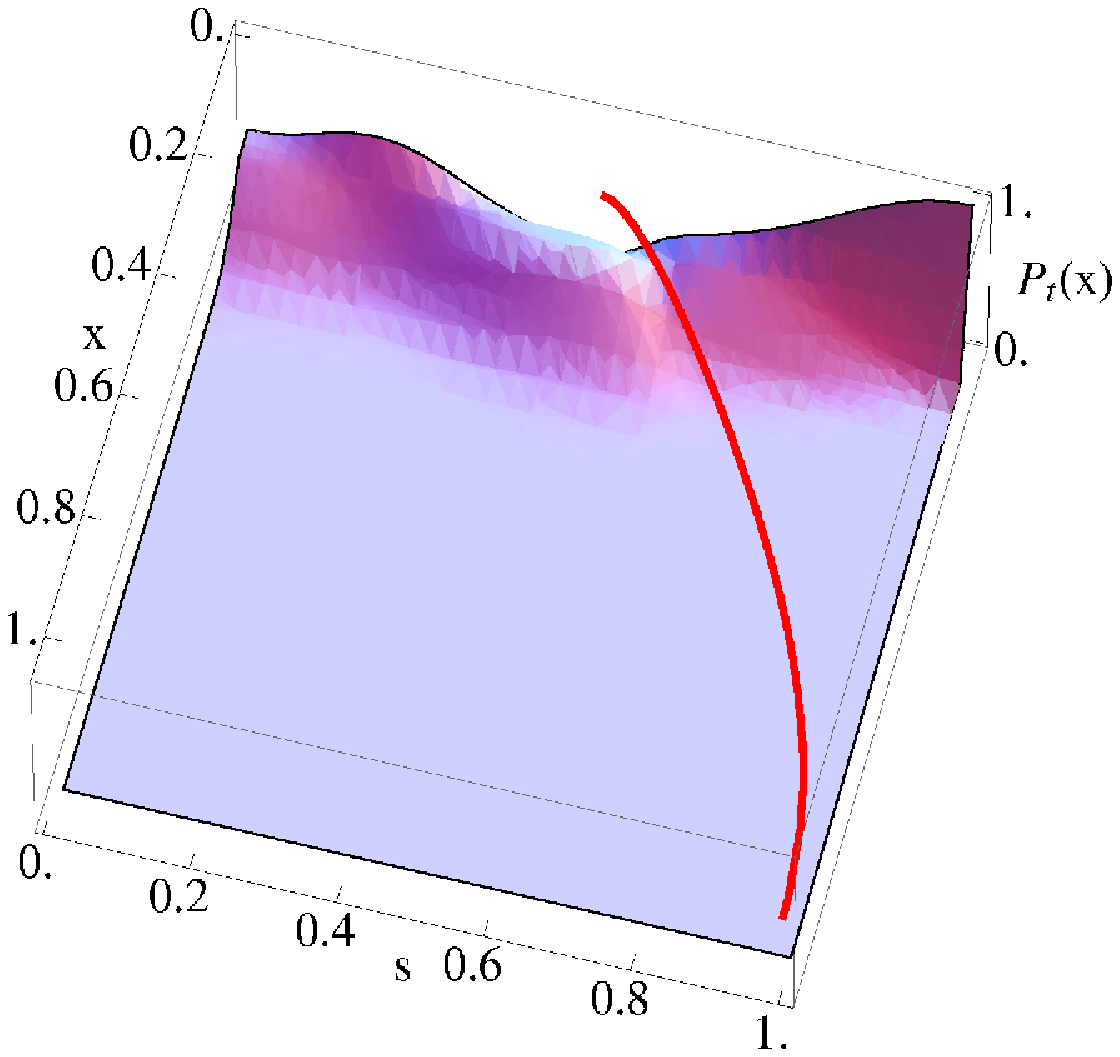} \\
   \includegraphics[height=5cm]{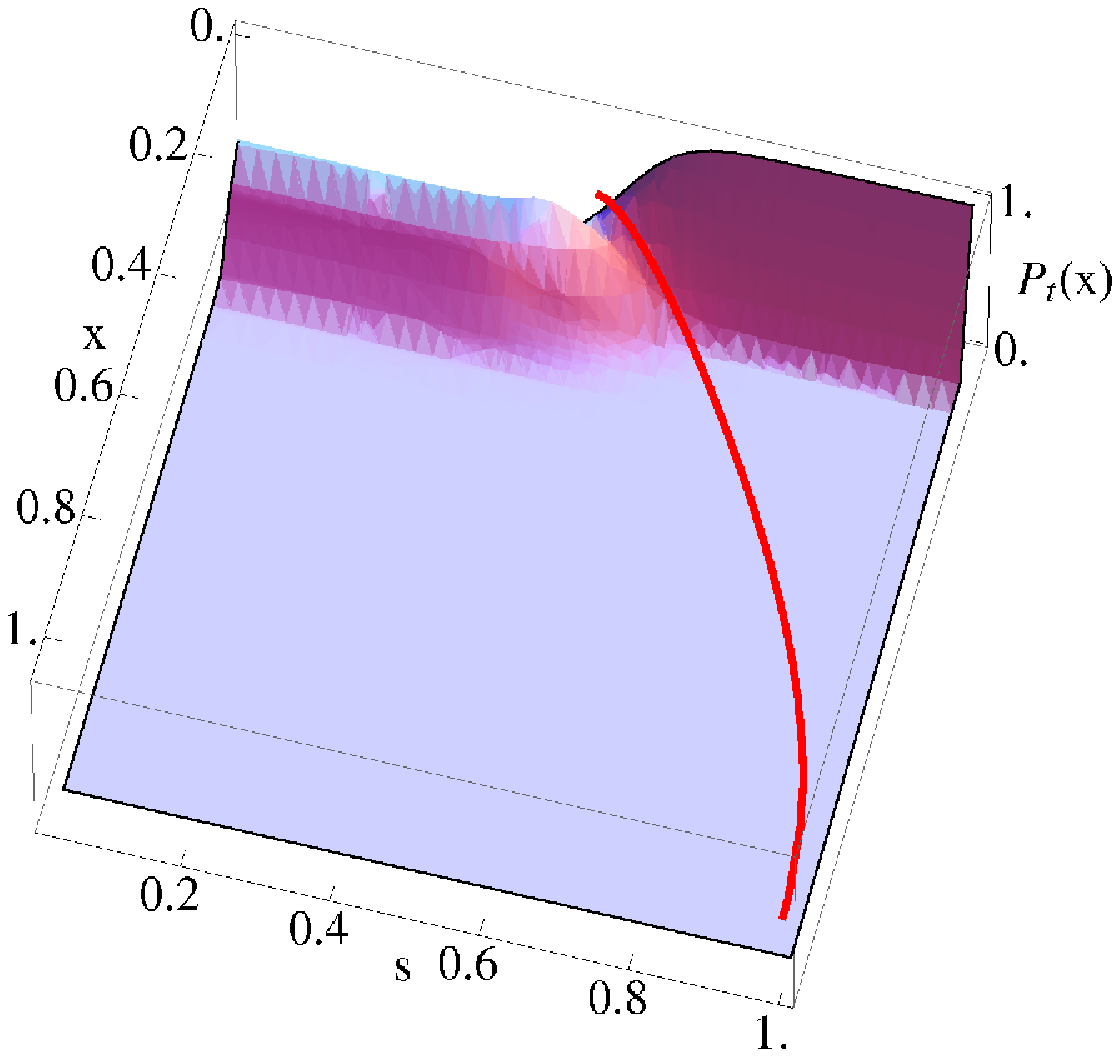}
   \includegraphics[height=5cm]{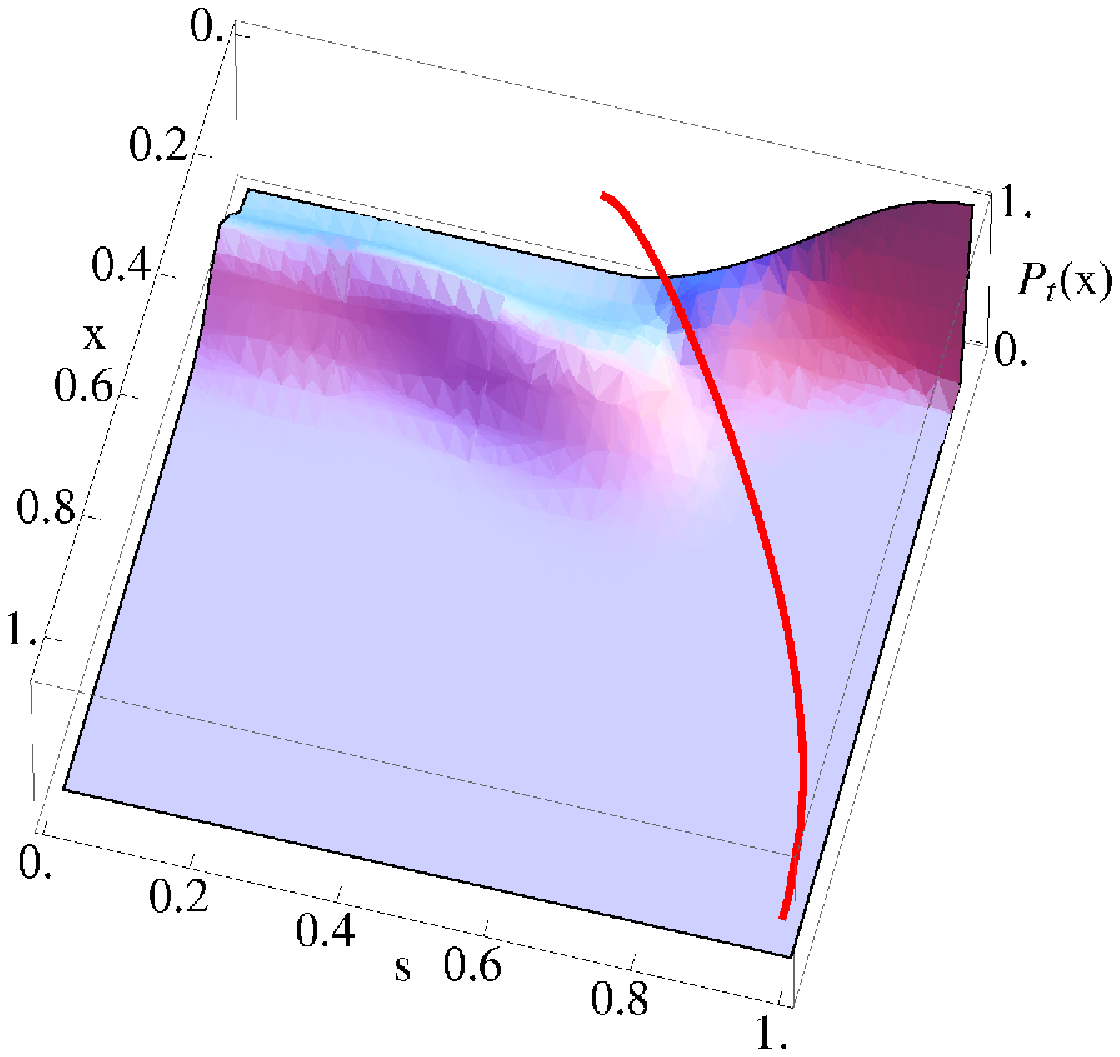}
    \caption{\label{fig:ME_inverse_evolution} 
   Levels population in the backward evolution case for a $20$ level system and $T=10$.  Phenomenological rate equation ($T~b=0.01$) (upper left), effective quantum chain with constant matrix element (upper right), effective chain with improved matrix elements (lower left), full numerical evolution (lower right).}
  \end{center}
\end{figure*}

Numerical solutions of the master equation (\ref{eq:classicalMQ}) are displayed in Fig.~\ref{fig:ME_direct_evolution} (up left).
This (very) simple model fairly reproduces some features of the computed evolutions. Indeed, for the forward evolution, we find the sequential transition driven by the CGC curve while for the backward evolution, a sudden transition to excited states and a subsequent saturation effect (after crossing the CGC) are recovered.

However, in the backward case for instance [ Fig.~\ref{fig:ME_inverse_evolution} - (up left) ], this simple model fails in describing correctly the lowest levels occupations. Indeed, the ground state remains here the most populated state during the evolution, a feature which is clearly not found in the full numerical simulations.

At this point an important remark must be done.
In this phenomenological model, we choose a transition rate (\ref{eq:tr_rate}) which depends on the inverse of the square of the energy gaps. This choice of an ``adiabatic''-like transition rate is, in a certain sense, arbitrary. Another possible choice would have been to take $\Gamma_{i,j}(s) \propto \exp{\{- b (\Delta^{i,j}(s))^2\}}$ (with $b\geq 0$), which mimics the Landau-Zener transition rate. We tried to plug this type of behavior in the rate equation, but could not find any reasonable agreement with the numerical results. This suggests that, for the present model, a generic Landau-Zener-like description is not appropriate.

\section{Simplified quantum model for the adiabatic evolution} 

In order to better describe the full dynamical process, we need to improve the previous approach, and incorporate 
quantum effects more precisely, as follows. If the evolving quantum state is written in the instantaneous basis of $H(s)$, 
with the gauge choice $\langle \dot{\eta}_n(s) | \eta_n(s) \rangle = 0$, the $a_m(s)$ coefficients satisfy the equation 

\begin{figure}[t]
  \begin{center}
   \includegraphics[height=5cm]{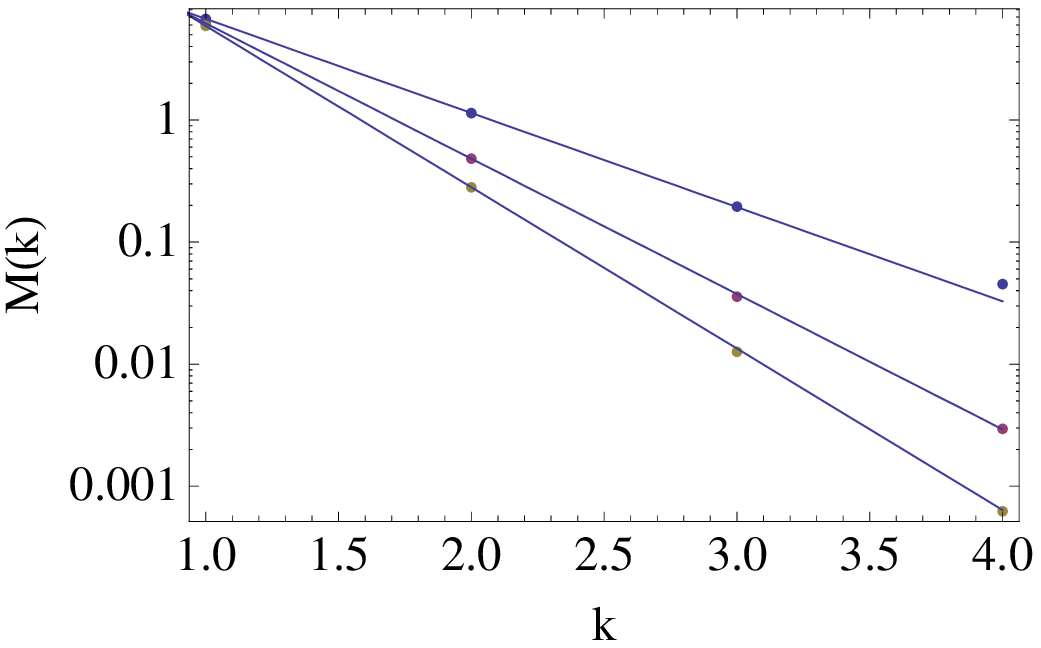}
    \caption{\label{fig:matrix_elements} Logarithmic plot of $M(k)= |\langle \eta_{n+k} (s)| \partial_s H(s) |\eta_n (s) \rangle|$ (with $n=10$ and $N=20$ ) as a function of the levels distance $k$ and for different $s$ (from lower to upper curves: $s=0.75$, $s=0.85$ and $s=0.90$), together with linear fits. }
  \end{center}
\end{figure}

\begin{equation}
 \partial_s a_m(s) = - i \epsilon_m(s) a_m(s)- \sum_{n \neq m} 
\frac{\langle \eta_m (s)| \partial_s H(s) |\eta_n (s) \rangle}{\Delta^{n,m}(s)} a_n(s)
\label{eq:quantumMEQ}
\end{equation}
where, as above, $\Delta^{n,m}(s)$ is the instantaneous gap between levels $n$ and $m$.
Note that solving equation (\ref{eq:quantumMEQ}) requires the knowledge of both the time-dependant eigenvalues and the matrix elements $\langle \eta_m (s)| \partial_s H(s) |\eta_n (s) \rangle$.

Let us further simplify this equation by using the numerically derived gaps and use an approximate form for the matrix elements. In addition, we limit the transitions from level $n$ to neighboring levels $n+1$ and level $n-1$, leading to an effective 1D quantum chain model.
As can be seen in figure \ref{fig:matrix_elements}, this is justified far from the CGC curve because matrix elements clearly display an exponentially decay form with distance in the effective chain.

We then use two forms for the matrix elements.
In the simplest case, we take them equal for any inter-level transition and constant with $s$. This already improves the previous phenomenological approach, as can be seen in figure \ref{fig:ME_direct_evolution} (up-right).
To further improve our model, we numerically compute the matrix elements in equation (\ref{eq:quantumMEQ}).
We find that in general, for a given level $n$,  $\langle \eta_n (s)| \partial_s H(s) |\eta_{n+1} (s) \rangle$
have a maximum near the CGC curve and then show a fast decrease. We fit this $s$-dependence with a Gaussian form, centered at the critical point. In addition we find that these matrix elements have {\it maximal} values, as a function of $n$, which can be well approximated by a logarithmic behavior.
We therefore write them as $\langle \eta_n (s)| \partial_s H(s) |\eta_{n+1} (s) \rangle = (a + b \log n) \exp{(-\gamma(s-s_0(n))^2)}$, where $a$, $b$ and $\gamma$ are fitted parameters and $s_0(n)$, the $s$ value where the gap between levels $n$ and $n+1$ is minimum, is very closely approximated from the analytic expression of the CGC curve.

The numerical simulation with this latter approximation is shown in figure \ref{fig:ME_direct_evolution} (down left).
It shows a clear improvement with respect to the phenomenological approach (\ref{eq:classicalMQ}), and even to the above constant matrix element approximation, in particular after crossing the CGC line, where the depleted population of the lowest levels is better reproduced. 

Backward evolution, treated with the same approximations, are presented in figure \ref{fig:ME_inverse_evolution}, with similar trends as in the forward case.

Let us stress that the  qualitative form of the computed evolution does not depends critically on the $s$ dependence of the matrix elements. Analogous results are obtained within a vast range of $\gamma$ coefficients and even with a
different functional dependence, as long as the approximating function remains well peaked around the critical point.
This confirms that the CGC drives the main feature of the dynamics; on the other hand, the absolute value of the matrix elements determine the ``fine details'' of the evolution, such as the ratio of the population levels.

A final remark concerns one important basic assumption of the above approaches (both phenomenological and quantum),  that this system is well approximated by an effective chain with only 
nearest neighbor transitions between levels.
This is true only {\it far from the CGC curve}; near the critical point, long ``distance'' transitions occur.
This point has been analyzed in a semi-classical framework, and will be presented elsewhere \cite{correlation_lenght}.

\section{Conclusion} 

We have studied time dependant dynamics of the Lipkin-Meshkov-Glick model driven across its Quantum Phase Transition Point. The dynamics of the quantum evolution, not restricted to the lowest level occupancy, is determined by the spectral critical gap curve, where the energy gaps vanish at the thermodynamic limit.
In order to compare with the full numerical solution, we have developed simplified models for the transitions during the evolution.

First, we use a phenomenological rate equation approach, with adiabatic-like transition rates, which already recovers the role of the CGC curve in driving the main quantum evolution. But this approach misses some important features of the quantum evolution.

We then improve our description by building a quantum model, which treats the inter-level jumps as the consequence of an effective interaction  between the instantaneous levels of the ``s''-dependant Hamiltonian, restricted to nearest-neighbor level interactions. In a first step, this interaction is only varied following the values of the ``s''-dependant gaps, which already compares better with the full numerical solution. We then further improve this effective chain model by including an approximate form for the rate of change of the Hamiltonian averaged over neighboring levels. In that case, the main features of the quantum evolution are recovered.

Future investigations should focus on a finer description of the quantum evolution near the critical curve. In particular, long range interactions between instantaneous levels, which come close in energy near that curve, have to be taken into account.

\acknowledgments
We are grateful to J. Vidal for fruitful and stimulating discussions. 
PR was partially supported by FCT and EU FEDER through POCTI and POCI, namely via QuantLog POCI/MAT/55796/2004 Project of CLC-DM-IST, SQIG-IT and grant SFRH/BD/16182/2004/2ZB5.
\\
\\
\textit{Note added - } While finishing this manuscript we became aware of a recent work by Caneva \textit{et al.} \cite{Caneva_2008} where the adiabatic dynamics for the LMG model is also studied. They focus on computing and analyzing the residual annealing energy as a function of the total adiabatic time. In particular, they numerically identify  three different regimes for the system evolution. For intermediate adiabatic times, the Kibble-Zurek mechanism, based on an analysis of ``defects'' and domain walls, used in \cite{Zurek_2005} for one-dimensional systems is invoked to analyze the computed behavior. We may just remark here that this may be questionable : the LMG model corresponds to a fully connected  set of spins $1/2$, therefore asymptotically in infinite dimension, and it is not clear how to adapt the Kibble-Zurek approach in that case.

\end{document}